\documentclass[showpacs,amssymb,twocolumn,aps,nofootinbib]{revtex4}
\usepackage{amsmath}
\usepackage{amstext}
\usepackage{amsopn}
\usepackage{amsfonts}
\usepackage{amssymb}
\usepackage{amssymb}
\usepackage{bbm}
\usepackage{accents}
\usepackage{empheq}%for overbracket
\usepackage{graphicx}
\usepackage{epsf}
\usepackage{graphics}
\usepackage{xcolor}
\usepackage[latin1]{inputenc}

\begin{document}

\title{Proper time method in de Sitter space}

\author{Ashok K. Das$^{a,b}$ and Pushpa Kalauni$^{c}$}
\affiliation{$^a$ Department of Physics and Astronomy, University of Rochester, Rochester, NY 14627-0171, USA}
\affiliation{$^b$ Saha Institute of Nuclear Physics, 1/AF Bidhannagar, Calcutta 700064, India}
\affiliation{$^{c}$ Instituto de Física, Universidade de São Paulo, 05508-090, São Paulo, SP, Brazil}

\begin{abstract}
We use the proper time formalism to study a (non-self-interacting) massive Klein-Gordon theory in the two dimensional de Sitter space. We determine the exact Green's function of the theory by solving the DeWitt-Schwinger equation as well as by calculating  the operator matrix element. We point out how the one parameter family of arbitrariness in the Green's function arises in this method.
\end{abstract}


\maketitle

\section{Introduction}
Schwinger's proper time method  \cite{schwinger} is a powerful method in quantum field theory in flat space-time. It introduces a gauge invariant regularization and leads to one loop effective actions  and  Green's functions (propagators) for field theories quite efficiently even at finite temperature \cite{farina, frenkel,frenkel1}. For example, let us consider the free, massive Klein-Gordon theory in flat space-time described by the Lagrangian density
\begin{equation}
{\cal L} = \frac{1}{2}\left(\partial^{\mu}\phi \partial_{\mu}\phi - m^{2}\phi^{2}\right) = \frac{1}{2}\phi \widehat{H} \phi,\label{1}
\end{equation}
where we have integrated by parts and have identified the operator
\begin{equation}
\widehat{H} = - \left(\Box + m^{2}\right),\qquad \Box = \eta^{\mu\nu}\partial_{\mu}\partial_{\nu}.\label{2}
\end{equation}
The (two point) Green's function of the theory is given by
\begin{equation}
\widehat{G} = \frac{1}{\widehat{H}},\qquad G(x,x') = \langle x|\widehat{G}|x'\rangle,\label{3}
\end{equation}
which can be written in the integral form
\begin{equation}
\widehat{G} = -i\int_{0}^{\infty} d\tau\,e^{i\tau \widehat{H}}.\label{4}
\end{equation}
Here the \lq\lq$i\epsilon$" prescription is understood in the exponent for convergence at the upper limit and $\tau$ is an auxiliary variable known as the \lq\lq proper time". In fact, we note that the operator $e^{i\tau\widehat{H}}$ can be thought of as the $\tau$ evolution operator (acting on bra states) with the generator given by $\widehat{H}$ which can be thought of as the Hamiltonian for $\tau$-translation so that we can write
\begin{align}
& K (x,x';\tau) = \langle x,\tau|x',0\rangle = \langle x|e^{i\tau \widehat{H}}|x'\rangle,\notag\\
& G(x,x') =-i\int_{0}^{\infty} d\tau\,\langle x,\tau|x',0\rangle = -i\int_{0}^{\infty} d\tau\, K(x,x';\tau).\label{5}
\end{align}
The function $K(x,x';\tau)$ is known as the heat kernel of the operator $\widehat{H}$ as will become clear shortly.

The Green's function $G(x,x')$ can be determined by solving the operator Heisenberg equations as demonstrated by Schwinger \cite{schwinger} (see also \cite{farina}). Here we discuss an alternate operator method which will be useful for our purpose. Identifying $\hat{p}_{\mu} = i\partial_{\mu}$, we can write the Hamiltonian in \eqref{2} as
\begin{equation}
\widehat{H} = \hat{p}^{2} - m^{2}.\label{6}
\end{equation}
As a result, introducing a complete momentum basis states in \eqref{5} in $D$ space-time dimensions, the heat kernel can be written as
\begin{align}
K(x,x';\tau) & = \int d^{D}p\,\langle x|p\rangle\langle p|e^{i\tau \widehat{H}}|x'\rangle\notag\\
& = e^{-i\tau m^{2}} \int \frac{d^{D}p}{(2\pi)^{D}}\, e^{i\tau p^{2} - ip\cdot (x-x')}.\label{7}
\end{align}
This can be trivially integrated to give
\begin{equation}
K(x,x';\tau) = \left(\frac{i}{4\pi}\right)^{\frac{D}{2}} \tau^{-\frac{D}{2}}\,e^{-im^{2}\tau - \frac{i(x-x')^{2}}{4\tau}}.\label{8}
\end{equation}
This has the expected behavior, namely, as $\tau\rightarrow 0$, we have $K(x,x';\tau)\rightarrow \delta^{D}(x-x')$. This heat kernel can be integrated over $\tau$ (see \eqref{5}) with proper ($i\epsilon$) regularizations to give the Feynman Green's function  \cite{GR}
\begin{align}
G(x-x') & = \left(\frac{1}{2\pi}\right)^{\!\frac{D}{2}}\!\!\!\!\left(\frac{(im)^{2}}{(x-x')^{2}-i\epsilon}\right)^{\!\frac{(D-2)}{4}}\notag\\
&\qquad \times K_{-\frac{D}{2}+1} (im\sqrt{(x-x')^{2}}),\label{9}
\end{align}
where $K_{\nu} (z)$ represents the Bessel function of the second kind. This result shows that the Green's function in flat space-time depends only on the invariant length or the geodesic distance between the two points.

When the theory is interacting with a gravitational background, the operator method becomes quite complicated, in general. In this case one uses a variation of the proper time method known as the DeWitt-Schwinger method \cite{DeWitt,bunchdavies,parker,barcelos}   which is also known as  the heat kernel method \cite{birrelldavies,fulling,vassilevich}. Here we note from \eqref{5} that we can write
\begin{equation}
-i\partial_{\tau} K(x,x';\tau) = H(x) K(x,x';\tau),\label{10}
\end{equation}
where $H(x)$ denotes the coordinate representation of the \lq\lq Hamiltonian" operator $\widehat{H}$. This equation is like the heat equation (or the Schr\"{o}dinger equation) which is why $K(x,x';\tau)$ is known as the heat kernel. In a general gravitational background, the operator $H(x)$ is complicated so that the heat kernel and, therefore, the Green's function cannot be obtained in a closed form. In this case, there are well known methods to determine the heat kernel as an asymptotic expansion (upto certain order) and then integrating over $\tau$ it is possible to determine an approximate form of the Green's function (see, for example, \cite{parker}). It turns out that the Green's function in a general gravitational background depends on both the coordinates independently and not only on the geodesic distance between the two points. 

One would hope that the proper time method (or the DeWitt-Schwinger method) would simplify in maximally symmetric spaces much like in flat space time. With this in mind, we would like to study this method and determine the Green's function for a massive Klein-Gordon field in de Sitter space (which is maximally symmetric). Furthermore, the Green's function, for a massive scalar field, has been determined from the operator methods in quantum field theory \cite{bunchdavies,gutzwiller,thirring,chernikov,tagirov,mottola,allen} and it is known that there is a (nonunique) one parameter family of vacua in de Sitter space leading to a nonunique Green's function \cite{mottola,allen}. Since the proper time method is  completely based on path integrals, it would be interesting to see how the one parameter ambiguity arises in this method. For simplicity we will restrict ourselves to de Sitter space in two dimensions, but the analysis can be trivially generalized to higher dimensions.

Our discussion is organized as follows. In section {\bf II}, we describe the global coordinates for the two dimensional de Sitter space and introduce the (non self-interacting) massive Klein-Gordon field theory in this metric background. The symmetry group of the two dimensional de Sitter space is the non-compact group $SO(2,1)$ and we discuss the coordinate representations for the generators of this group as well as the representations of this group in section {\bf III}. In section {\bf IV}, we determine the exact Green's function of this theory using the DeWitt-Schwinger method and point out how the one parameter family of arbitrariness arises in this method. We conclude with a brief summary in section {\bf V}  and discuss the (proper time) operator method of determining the Green's function in the  appendix. 

\section{de Sitter space in two dimensions}

The two dimensional de Sitter space \cite{thirring,strominger,kim,moschella,akhmedov} can be embedded in a three dimensional space satisfying
\begin{equation}
(X^{0})^{2} - (X^{1})^{2} - (X^{2})^{2} = - R^{2},\label{11}
\end{equation}
where $X^{0}, X^{1}, X^{2}$ denote the cordinates of the three dimensional space and $R$ is a constant which we set to unity for simplicity ($R=1$). We use the Bjorken-Drell metric which is diagonal with the signatures $(+,-,-)$. The two dimensional coordinates, $(t,\theta)$ of the de Sitter space can now be defined through the relations
\begin{align}
X^{0} & = \sinh t,\notag\\
X^{1} & = \cosh t\,\cos\theta,\notag\\
X^{2} & = \cosh t\,\sin\theta,\label{12}
\end{align}
where $-\infty < t < \infty, 0\leq \theta \leq 2\pi$ (remember $R=1$). These are known as global coordinates of the two dimensional de Sitter space. The invariant line element can now be easily determined to be
\begin{equation}
ds^{2} = dt^{2} - \cosh^{2} t\,d\theta^{2},\label{13}
\end{equation}
leading to
\begin{equation}
g_{tt} = 1,\quad g_{\theta\theta} = -\cosh^{2} t,\quad \sqrt{-g} = \cosh t.\label{14}
\end{equation}

A (non self-interacting) massive Klein-Gordon theory in this space is described by the action 
\begin{align}
S = \int d^{2}x\,\sqrt{-g}\,{\cal L} & = \frac{1}{2}\int d^{2}x\,\sqrt{-g}\left(g^{\mu\nu} \partial_{\mu}\phi \partial_{\nu}\phi - m^{2}\phi^{2}\right)\notag\\
& = \frac{1}{2} \int d^{2}x\,\sqrt{-g}\,\phi \widehat{H}\phi,\label{15}
\end{align}
where we have integrated by parts and have identified (see \eqref{2})
\begin{equation}
\widehat{H} = -\left(\Box+m^{2}\right) = -\frac{1}{\sqrt{-g}}\partial_{\mu}\sqrt{-g}g^{\mu\nu}\partial_{\nu} - m^{2}.\label{16}
\end{equation}
Here we are using the notation $x^{\mu} = (t,\theta)$ with $\mu=t,\theta$. It is worth pointing out here that the \lq\lq Hamiltonian", $\widehat{H}$, is not Hermitian with respect to the conventional integration measure $\int d^{2}x$, but is Hermitian with respect to the integration measure $\int d^{2}x\,\sqrt{-g}$. This, in turn, induces the inner product of the coordinate states to be defined with a factor of $\frac{1}{\sqrt{-g}}$ (see also, for example, \cite{parker}). The D' Alembertian, in this two dimensional de Sitter space has the explicit form
\begin{equation}
\Box = \partial_{t}^{2} + \tanh t\,\partial_{t} - \text{sech}^{2} t\,\partial_{\theta}^{2}.\label{17}
\end{equation}

We can now follow the discussion in \eqref{3}-\eqref{5} and determine the Green's function in terms of the heat kernel. We will do so by solving the DeWitt-Schwinger equation \eqref{10} in section {\bf IV} leaving the operator determination to the appendix. However, let us first discuss the structure of the symmetry group of the theory in the next section.

\section{$\mathbf{SO(2,1)}$ and its representations}

The maximal symmetry group of the two dimensional de Sitter space is given by $SO(2,1)$. This is easily seen from the fact that the invariant length in \eqref{11} remains unchanged under an arbitrary $SO(2,1)$ transformation. This will reflect in the fact that $\widehat{H}$ coming from the massive Klein-Gordon theory will also be $SO(2,1)$ invariant. Let us see this explicitly by constructing the infinitesimal generators of $SO(2,1)$. We note that this is a non-compact group so that finite dimensional representations can not be unitary. Therefore, we concentrate only on infinite dimensional representations which will be unitary. Let us look at the coordinate representations of the three generators of the group which can be written in terms of the two dimensional de Sitter coordinates $(t,\theta)$ as
\begin{align}
J_{1} & = i\left(\cos\theta\,\partial_{t} - \tanh t\,\sin\theta\,\partial_{\theta}\right),\notag\\
J_{2} & = i\left(\sin\theta\,\partial_{t} + \tanh t\,\cos\theta\,\partial_{\theta}\right),\notag\\
J_{3} & = -i\partial_{\theta}.\label{18}
\end{align}
The generators can be seen to be Hermitian with respect to the integration measure $\int d^{2}x\,\sqrt{-g}$ as pointed out  after eq. \eqref{16}. It can now be directly checked that they satisfy the Lie algebra
\begin{align}
\left[J_{1}, J_{2}\right] & = - i J_{3},\notag\\
\left[J_{2}, J_{3}\right] & = i J_{1},\notag\\
\left[J_{3}, J_{1}\right] & = i J_{2}.\label{19}
\end{align}
This is similar to the Lie algebra of $so(3)$ except for the sign difference in the first commutator which makes it the non-compact $so(2,1)$ Lie algebra. (This algebra is also isomorphic to $su(1,1)$.) We note that $J_{3}$ corresponds to the generator of rotations in the $X^{1}$-$X^{2}$ plane and is, therefore, a compact generator.

For future use, we note here that if we define
\begin{equation}
z = i \sinh t,\label{19a}
\end{equation}
the generators can also be written as
\begin{align}
J_{1} & = - \frac{1}{\sqrt{1-z^{2}}}\left((1-z^{2})\,\cos\theta\,\partial_{z} + z \sin\theta\,\partial_{\theta}\right),\notag\\
J_{2} & = - \frac{1}{\sqrt{1-z^{2}}}\left((1-z^{2})\,\sin\theta\,\partial_{z} - z \cos\theta\,\partial_{\theta}\right),\notag\\
J_{3} & = - i\partial_{\theta}.\label{19b}
\end{align}

The quadartic Casimir of the algebra can now be constructed and has the coordinate representation
\begin{equation}
J^{2} =  J_{1}^{2} + J_{2}^{2} -   J_{3}^{2} =  - \Box,\label{20}
\end{equation}
where the D' Alembertian in the de Sitter space is defined in \eqref{17}. Since all the generators $J_{i}, i=1,2,3$ commute with the Casimir operator $J^{2}$, it is clear that all the generators also commute with $\widehat{H}$ defined in \eqref{16}. Therefore, $SO(2,1)$ defines the symmetry group of the \lq\lq Hamiltonian" associated with the $\tau$-evolution and the eigenstates of $\widehat{H}$ will be described by the angular momentum states (representations) of $SO(2,1)$.

Unitary representations of $SO(2,1)$ are well studied in the literature  \cite{thirring,bargmann,naimark,vilenkin,ruhl,wybourne,gursey,maharana} and we will only summarize the results here. If we choose to diagonalize the generator $J_{3}$, its eigenvalues will take integer values $k=0,\pm 1, \pm 2,\cdots$ (in this scalar example) since the $\theta$ angle is compact with a range of $2\pi$. Representations are then completely determined by the eigenvalues $\nu$ of the Casimir operator $J^{2}$. (Conventionally the eigenvalues of $J^{2}$ are defined as $-\nu(\nu+1)$ and it is clear that they can take indefinite values). There are two infinite dimensional discrete series denoted by $D_{\nu}^{\pm}$ where the eigenvalues $k$ are bounded from below/above. In this case, $\nu$ takes negative integer values (in this example of a scalar theory) and the qudratic Casimir has negative semi-definite eigenvalues. However, the representation of interest to us is known as the continuous series (also known as the principal series) where the eigenvalue $\nu$ takes a continuum of complex values $\nu = -\frac{1}{2} + i\lambda,\, \lambda > 0^{+}$ such that the quadratic Casimir operator has a continuum of positive definite eigenvalues. (There is another representation known as the supplementary series/exceptional series which is not very relevant in the study of physical systems.) 

For the continuous/principal series, the normalized eigenfunctions (spherical harmonics) are given by  \cite{vilenkin,maharana,magnus,jackiw,grosche,virchenko}
\begin{equation}
Y_{-\frac{1}{2}+i\lambda}^{k} (x) = \frac{\Gamma(\frac{1}{2}-k+i\lambda)}{\Gamma(i\lambda)} P_{-\frac{1}{2}+i\lambda}^{k} (t) \frac{e^{ik\theta}}{\sqrt{2\pi}},\label{21}
\end{equation}
and satisfy the orthonormality and completeness relations 
\begin{align}
& \int d^{2}x\,(Y_{-\frac{1}{2}+i\lambda}^{k}(x))^{*} Y_{-\frac{1}{2}+i\lambda'}^{k'} (x) = \delta_{\lambda\lambda'} \delta_{kk'},\notag\\
& \int_{0}^{\infty}d\lambda \sum_{k} Y_{-\frac{1}{2}+i\lambda}^{k}(x) (Y_{-\frac{1}{2}+i\lambda}^{k}(x'))^{*} = \delta^{2}(x-x').\label{22}
\end{align}
The completeness relation, in particular, will be very useful in determining the Green's function in this two dimensional de Sitter space.

\section{DeWitt-Schwinger method}

In this section, we will determine the exact Green's function for the massive Klein-Gordon theory in the two dimensional de Sitter space using the DeWitt-Schwinger equation \eqref{10} (as well as \eqref{5})
\begin{equation}
-i\partial_{\tau} K (x,x';\tau) = H(x) K (x,x';\tau),\label{23}
\end{equation}
where (see \eqref{16}-\eqref{17})
\begin{equation}
H(x) = -\left(\partial_{t}^{2} + \tanh t\,\partial_{t} - \text{sech}^{2} t\,\partial_{\theta}^{2} + m^{2}\right).\label{24}
\end{equation}
The solution of \eqref{23} can be written in the factorizable form
\begin{equation}
K(x,x';\tau) = \Theta_{k} (\theta') F(t,t')\, e^{i\tau E}\,\frac{e^{ik\theta}}{\sqrt{2\pi}},\label{25}
\end{equation}
where $E$ is an arbitrary separation constant, $k$ takes integer values (for the solution to be single valued) and $F(t,t')$ satisfies ($\Theta(\theta')$ is a function to be determined)
\begin{equation}
\left(\partial_{t}^{2} + \tanh t\,\partial_{t} + (E+m^{2}) + \frac{k^{2}}{\cosh^{2} t}\right) F(t,t') = 0.\label{26}
\end{equation}

If we define a new variable
\begin{equation}
z = i \sinh t,\label{27}
\end{equation}
then \eqref{26} can be rewritten as
\begin{equation}
\left(\frac{d}{dz} (1-z^{2}) \frac{d}{dz} - (E+m^{2}) - \frac{k^{2}}{(1-z^{2})}\right)F(z,z') = 0.\label{28}
\end{equation}
This is the equation for the associated Legendre function $P_{-\frac{1}{2}+i\lambda}^{k} (z)$ (corresponding to the continuous representation of $SO(2,1)$ discussed in the previous section) with the identification
\begin{align}
& E+m^{2} = \lambda^{2}+\frac{1}{4},\notag\\
& E = \lambda^{2} - s^{2},\qquad s^{2} = m^{2} - \frac{1}{4}.\label{29}
\end{align}
Here we assume that $m^{2} > \frac{1}{4}$. (The inequality should really come as $(mR)^{2} > \frac{1}{4}$, but we have set $R$ to unity.) Therefore, the solution of \eqref{28} can be written (with some normalization factor, see \eqref{21}) as
\begin{equation}
F(z,z') = f_{E,k}(z') \frac{\Gamma (\frac{1}{2}-k+i\lambda)}{\Gamma(i\lambda)}\,P_{-\frac{1}{2}+i\lambda}^{k} (z),\label{30}
\end{equation}
with $E=\lambda^{2}-s^{2}$ as determined in \eqref{29}.

As a result, the general solution of the DeWitt-Schwinger equation \eqref{23}-\eqref{24} (or the heat kernel) can be written as
\begin{align}
& K(x,x';\tau) = \langle x,\tau|x',0\rangle\notag\\
& = \int_{0}^{\infty}d\lambda\sum_{k}f_{\lambda,k}(z')\Theta_{k}(\theta') Y_{-\frac{1}{2}+i\lambda}^{k} (z,\theta)e^{i\tau(\lambda^{2}-s^{2})}.\label{31}
\end{align}
As $\tau\rightarrow 0$, this should be proportional to the delta function (in this limit, the heat kernel is simply the inner product of two coordinate states at $\tau=0$) and this determines the functions $f_{\lambda,k}(z')\Theta_{k}(\theta')$ so that we can write the heat kernel as
\begin{align}
K(x,x';\tau) & = \int_{0}^{\infty}d\lambda\sum_{k} Y_{-\frac{1}{2}+i\lambda}^{k}(z,\theta)(Y_{-\frac{1}{2}+i\lambda}^{k}(z',\theta'))^{*}\notag\\
&\qquad \times e^{i\tau (\lambda^{2}-s^{2})}.\label{32}
\end{align}
It can be checked now that
\begin{align}
& K(x,x';\tau) = \langle x,\tau|x',0\rangle\notag\\
&\xrightarrow{\tau\rightarrow 0} \delta(z-z')\delta (\theta-\theta') = \frac{\delta^{2}(x-x')}{\sqrt{-g}},\label{33}
\end{align}
as we would expect (see the discussion after eq. \eqref{16} as well as \cite{parker}). Here we have used the completeness relation of the spherical harmonics given in \eqref{22}.

The Green's function can now be obtained (see \eqref{5}) by integrating the heat kernel over $\tau$ (with proper $i\epsilon$ prescription) and leads to
\begin{equation}
G(x,x') = \int_{0}^{\infty} d\lambda\sum_{k} \frac{Y_{-\frac{1}{2}+i\lambda}^{k}(z,\theta)(Y_{-\frac{1}{2}+i\lambda}^{k}(z',\theta'))^{*}}{\lambda^{2}-s^{2}+i\epsilon},\label{34}
\end{equation}
which is easily checked to give the correct spectral representation of the Green's function. Using known gamma function identities \cite{GR,AS}
\begin{align}
& \Gamma(x)\Gamma(1-x) = \frac{\pi}{\sin\pi x},\notag\\
& \Gamma(\frac{1}{2}+x)\Gamma(\frac{1}{2}-x) = \frac{\pi}{\cos\pi x},\label{35}
\end{align}
as well as the well known relation for the associated Legendre functions
\begin{equation}
P_{-\frac{1}{2}+i\lambda}^{k}(z) = \frac{\Gamma(\frac{1}{2}+k+i\lambda)}{\Gamma(\frac{1}{2}-k+i\lambda)}\,P_{-\frac{1}{2}+i\lambda}^{-k}(z),\label{36}
\end{equation}
we can write
\begin{widetext}
\begin{align}
\sum_{k} Y_{-\frac{1}{2}+i\lambda}^{k}(z,\theta)(Y_{-\frac{1}{2}+i\lambda}^{k}(z',\theta'))^{*} & = \frac{\lambda}{2\pi}\frac{\sinh\pi\lambda}{\cosh\pi\lambda}\Big[P_{-\frac{1}{2}+i\lambda}(i\sinh t)P_{-\frac{1}{2}+i\lambda}(-i\sinh t')\notag\\
&\qquad + 2\sum_{k=1}^{\infty} (-1)^{k} P_{-\frac{1}{2}+i\lambda}^{k}(i\sinh t)P_{-\frac{1}{2}+i\lambda}^{-k}(-i\sinh t') \cos k(\theta-\theta')\Big]\notag\\
& = \frac{\lambda}{2\pi}\,\frac{\sinh\pi\lambda}{\cosh\pi\lambda}\, P_{-\frac{1}{2}+i\lambda} (Z).\label{37}
\end{align}
\end{widetext}
Here we have identified
\begin{equation}
Z = \sinh t \sinh t' - \cosh t \cosh t' \cos(\theta-\theta'),\label{38}
\end{equation}
which corresponds to the negative of the geodesic distance between $x$ and $x'$ and we have also used the addition theorem \cite{bateman} for the associated Legendre functions in the last step in \eqref{37}. Substituting \eqref{37} into \eqref{34} and recognizing that the integrand is an even function of $\lambda$, the contour integral determines the exact Green's function to be
\begin{equation}
G(x,x') = G(Z) = -\frac{i}{4}\,\frac{\sinh \pi s}{\cosh \pi s}\,P_{-\frac{1}{2}+is} (Z).\label{39}
\end{equation}
This can be compared with the results obtained from a direct analysis of the quantum field theory \cite{mottola,allen, bousso,barata} and agrees  except for the normalization factor $\frac{1}{\cosh\pi s}$ which is really necessary for the correct spectral representation of the Green's function to hold, as also noted elsewhere \cite{jackiw}.

It is clear from the definition of the geodesic distance in \eqref{38} that $Z$ is invariant under the $t$ coordinate redefinitions
\begin{equation}
 t\rightarrow t + i\pi,\quad t'\rightarrow t' - i\pi,\label{40}
\end{equation}
so that the Green's function has a periodicity of $2\pi$ in the time argument. This suggests \cite{das} that the Green's function in de Sitter space has a thermal character \cite{mottola,thirring1,boer} with a temperature $\frac{1}{2\pi}$. 

The one parameter family  \cite{mottola,allen,bousso,brunetti,mottola1}  of arbitrariness in the Green's function can also be seen in this derivation as follows. Let us recall that, in the quantum field theory approach, there is a nonuniqueness in the choice of the basis functions in which one expands the field operator. In fact, a general Bogoliubov transformation of the form ($\phi_{k} (\phi^{*}_{k})$ are coefficient functions of the annihilation (creation) operators respectively)
\begin{equation}
\begin{pmatrix}
\phi_{k}\\
\phi^{*}_{k}
\end{pmatrix} \rightarrow \begin{pmatrix}
\tilde{\phi}_{k}\\
\tilde{\phi}^{*}_{k}
\end{pmatrix} = U\begin{pmatrix}
\phi_{k}\\
\phi^{*}_{k}
\end{pmatrix},\label{40a}
\end{equation}
where 
\begin{equation}
U = \begin{pmatrix}
\cosh\alpha & \sinh\alpha\,e^{i\beta}\\
\sinh\alpha\,e^{-i\beta} & \cosh\alpha
\end{pmatrix},\label{41}
\end{equation}
with $\alpha,\beta$ two real parameters of the transformation, takes us to another set of equivalent basis functions. The matrices $U$ belong to the group $SU(1,1)$ and satisfy
\begin{equation}
U \sigma_{3} U^{\dagger} = \sigma_{3},\label{41a}
\end{equation}
where $\sigma_{3}$ denotes the third Pauli matrix. This nonuniqueness in the choice of the basis functions leads to a family of quantum vacua depending, in general, on two parameters $\alpha, \beta$. However, it is well known by now \cite{allen, bousso, Einhorn} that when the parameter $\beta \neq 0$, the vacua violate $CPT$ invariance. Therefore, one chooses $\beta=0$ which leads to an one parameter family of vacua and, therefore, Green's functions depending on the value of $\alpha$.

With the choice $\beta=0$, the $2\times 2$ transformation matrix in \eqref{41} becomes real and corresponds to a transformation matrix belonging to $SO(2,1)$
\begin{equation}
M = \begin{pmatrix}
\cosh\alpha & \sinh\alpha\\
\sinh\alpha & \cosh\alpha
\end{pmatrix},\label{41b}
\end{equation}
which satisfies
\begin{equation}
M \sigma_{3}M^{T}  = \sigma_{3}.\label{41c}
\end{equation}
We note that since $\widehat{H}$ is essentially the Casimir operator $J^{2}$ of $SO(2,1)$ (see, for example, \eqref{16} and \eqref{20}), it (as well as the differential equation) is invariant under $SO(2,1)$ transformations. The associated Legendre equation in \eqref{28} is a second order equation and, therefore, there are two linearly independent solutions which can be represented by $P_{-\frac{1}{2}+i\lambda}^{k} (z)$ and $P_{-\frac{1}{2}+i\lambda}^{k} (-z)$. Under a finite $SO(2,1)$ transformations, the two linearly independent solutions transform as
\begin{equation}
\begin{pmatrix}
P_{-\frac{1}{2}+i\lambda}^{k} (z)\\
P_{-\frac{1}{2}+i\lambda}^{k} (-z)
\end{pmatrix}\rightarrow \begin{pmatrix}
\cosh\alpha & \sinh\alpha\\
\sinh\alpha & \cosh\alpha
\end{pmatrix}\begin{pmatrix}
P_{-\frac{1}{2}+i\lambda}^{k} (z)\\
P_{-\frac{1}{2}+i\lambda}^{k} (-z)
\end{pmatrix}.\label{41d}
\end{equation}
In other words, under such a rotation
\begin{equation}
P_{-\frac{1}{2}+i\lambda}^{k}(z)\rightarrow \cosh\alpha\, P_{-\frac{1}{2}+i\lambda}^{k}(z) + \sinh\alpha\,P_{-\frac{1}{2}+i\lambda}^{k}(-z).\label{42}
\end{equation}
Using these as the new (rotated) solutions in \eqref{30}, we can calculate the Green's function as discussed in \eqref{34}-\eqref{37} which leads to
\begin{align}
& G(x,x') = G(Z) \rightarrow -\frac{i}{4}\,\tanh \pi s\notag\\
&\times\left(\cosh 2\alpha P_{-\frac{1}{2}+is}(Z) + \sinh 2\alpha\,P_{-\frac{1}{2}+is}(-Z)\right)\notag\\
&\quad = \cosh 2\alpha\, G(Z) +\sinh 2\alpha\,G(-Z).\label{43}
\end{align}
This leads to the well known one parameter family of arbitrariness in the Green's function in the proper time method.

\section{Conclusion}

In this paper we have extended the proper time method to de Sitter space. In particular, we have calculated the Green's function for the (non self-interacting) massive scalar field in two dimensional de Sitter space both by solving the DeWitt-Schwinger equation as well as by directly calculating the matrix element of the \lq\lq Hamiltonian" operator (see the appendix). This matches with the results derived earlier by studying the scalar field theory in de Sitter space except for an overall normalization constant which is necessary for the spectral representation of the Green's function to hold. We have pointed out the periodicity in the Green's function as well as how the one parameter family of arbitrariness arises in this method. 

To conclude, we note that, although our calculation was done in two dimensional de Sitter space for simplicity, the method generalizes to any higher dimensional de Sitter space in a natural manner. Let us discuss this briefly without going into technical details. The first thing to note is that in a $d$-dimensional de Sitter space the D' Alembertian continues to be given (up to a possible sign) by the quadratic Casimir operator, $J^{2}$, of $SO(d,1)$ (see, for example, \eqref{20}). Furthermore, in the global coordinates, one can write the D' Alembertian explicitly as
\begin{equation}
\Box = \frac{\partial^{2}}{\partial t^{2}} + (d-1)\tanh t \frac{\partial}{\partial t} + \frac{L^{2}}{\cosh^{2} t},\label{44}
\end{equation}
where $L^{2}$ denotes the quadratic Casimir operator of $SO(d-1)$ and depends only on the $(d-1)$ angular variables. Namely, $L^{2}$ corresponds to the angular part of the D' Aembertian (see, for example, \eqref{17} and we point out that the generators of angular momentum are defined with a factor of $i$). $L^{2}$ satisfies the eigenvalue equation
\begin{equation}
L^{2} Y(\Omega) = \ell (\ell + d-2) Y (\Omega),\label{45}
\end{equation}
where $\ell$ denotes the eigenvalues of the angular momentum operator, $Y$ the spherical harmonics and $\Omega$ denotes collectively the $(d-1)$ angular coordinates.

Thus, the Klein-Gordon equation or the DeWitt-Schwinger equation will be separable as discussed in section {\bf IV}. For example, we can write a solution of the free massive Klein-Gordon equation in the separable form
\begin{equation}
\phi(x) = \phi (t) Y(\Omega),\label{46}
\end{equation}
where the function $\phi(t)$ satisfies the equation
\begin{equation}
\left(\frac{d^{2}}{d t^{2}} + (d-1)\tanh t \frac{d}{d t} +m^{2} + \frac{\ell(\ell +d-2)}{\cosh^{2}t}\right)\phi(t)=0.\label{47}
\end{equation}
Redefining the variable as well as the function as \cite{pushpa}
\begin{equation}
z= i \sinh t,\quad \phi(z) = (1-z^{2})^{\frac{2-d}{4}}\,f(z),\label{48}
\end{equation}
equation \eqref{47} takes the form
\begin{equation}
\frac{d}{d z} (1-z^{2}) \frac{df(z)}{d z} - \left[m^{2}-\frac{d(d-2)}{4} + \frac{(2\ell + d-2)^{2}}{4(1-z^{2})}\right]f(z) = 0,\label{49}
\end{equation}
which corresponds to the associated Legendre equation. The associated Legendre functions solving this equation are given by $P_{\nu}^{\mu} (z)$ where
\begin{align}
\nu & = - \frac{1}{2} \pm i\sqrt{m^{2} - \frac{(d-1)^{2}}{4}},\notag\\
\mu & = \frac{1}{2}\left(2\ell + d-2\right).\label{50}
\end{align}
The DeWitt-Schwinger analysis of section {\bf IV} can now be carried out systematically.

\bigskip

\noindent{\bf Acknowledgments}
\bigskip

One of us (A. D.) would like to acknowledge helpful discussions with Profs J. Maharana and S. Panda. He would also like to thank the Departamento de F\'{i}sica Matem\'{a}tica in USP as well as the Institute of Physics for hospitality where this work was done. P. K. was supported in part by FAPESP (Brazil) - Process No. 2013/08090-9.
 
\appendix

\section{Green's function from an operator method}

In the main text, we have determined the Green's function for the massive Klein-Gordon theory in two dimensional de Sitter space by solving the DeWitt-Schwinger equation. Here we will briefly indicate how this can also be derived directly from the operator method discussed in \eqref{7}-\eqref{9}. Let us recall from \eqref{16} and \eqref{20} that in this case we can write
\begin{equation}
\widehat{H} = (J^{2} - m^{2}).\label{app1}
\end{equation}
Therefore, we see from \eqref{5} that we can write
\begin{equation}
K(x,x';\tau) = \langle x|e^{i\tau (J^{2}-m^{2})}|x'\rangle.\label{app2}
\end{equation}
Since the operator in the exponent is the quadratic Casimir of the group $SO(2,1)$, we can easily calculate its matrix element by inserting a complete basis of the angular momentum states in \eqref{app2}. Furthermore, following the discussion of the relevant representation of $SO(2,1)$ in this case, we obtain
\begin{widetext}
\begin{equation}
K (x,x';\tau) = \int_{0}^{\infty} d\lambda\sum_{k} \langle x|-\frac{1}{2}+i\lambda, k\rangle\langle -\frac{1}{2}+i\lambda,k|e^{i\tau (J^{2}-m^{2})}|x'\rangle  = \int_{0}^{\infty} d\lambda\sum_{k} Y_{-\frac{1}{2}+i\lambda}^{k}(z,\theta)(Y_{-\frac{1}{2}+i\lambda}^{k} (z',\theta'))^{*} e^{i\tau(\lambda^{2}-s^{2})},\label{app3}
\end{equation}
\end{widetext}
where $s$ is defined in \eqref{29} and 
\begin{equation}
\langle x|-\frac{1}{2}+i\lambda,k\rangle = Y_{-\frac{1}{2}+i\lambda}^{k}(z,\theta),\label{app4}
\end{equation}
denote the normalized spherical harmonics defined earlier. Expression \eqref{app3} can be compared with the heat kernel determined earlier in \eqref{32} and the Green's function can now be derived completely parallel to the steps discussed in section {\bf IV}.

\end{document}